\documentclass[twocolumn,showpacs,showkeys,amsmath,amssymb,pra]{revtex4}

\usepackage{graphicx}   
\usepackage{dcolumn}    
\usepackage{bm}         
\begin{document}

\title{Photon-assisted tunneling in optical lattices: \\ Ballistic transport of interacting boson pairs}

\author{Christoph Weiss}
\email{weiss@theorie.physik.uni-oldenburg.de}

\affiliation{Institut f\"ur Physik, Carl von Ossietzky Universit\"at,
                D-26111 Oldenburg, Germany
}
\affiliation{Laboratoire Kastler Brossel, \'Ecole Normale Sup\'erieure,
Universit\'e Pierre et Marie-Curie-Paris 6, 24 rue Lhomond, CNRS,
                F-75231 Paris Cedex 05, France
}

\author{Heinz-Peter Breuer}

\affiliation{Physikalisches Institut, Albert-Ludwigs-Universit\"at
Freiburg, Hermann-Herder Strasse 3, D-79104 Freiburg, Germany}


\date{\today}

\begin{abstract}
In a recent experiment [PRL 100, 040404 (2008)] an analog of
photon-assisted tunneling has been observed for a Bose-Einstein
condensate in an optical lattice subject to a constant force plus
a sinusoidal shaking. Contrary to previous theoretical
predictions, the width of the condensate was measured to be
proportional to the square of the effective tunneling matrix
element, rather than a linear dependence. For a simple model of
two \emph{interacting}\/ bosons in a one-dimensional optical
lattice, both analytical and numerical calculations indicate that
such a transition from a linear to a quadratic dependence can be
interpreted through the ballistic transport and the corresponding
exact dispersion relation of bound boson pairs.
\end{abstract}

\maketitle


\section{Introduction}
Bose-Einstein condensates (BECs) in optical lattices provide an
excellent tool to study solid state
systems~\cite{JakschZoller05}. 
One of the methods which are currently established
experimentally~\nocite{LignierEtAl07}\cite{LignierEtAl07,SiasEtAl08,Zenesini08}
for BECs in an optical lattice is tunneling control via
time-periodic potential
differences~\nocite{GrossmannEtAl91,Holthaus92b,GrifoniHanggi98}\cite{GrossmannEtAl91,Holthaus92b,GrifoniHanggi98,EckardtEtAl05}.
Effects investigated theoretically in periodically shaken systems
include multi-particle
entanglement~\nocite{Creffield2007}\cite{Creffield2007,TeichmannWeiss07}
and nonlinear Landau-Zener processes~\cite{ZhangEtAl08}.

The experimental
realization~\nocite{LignierEtAl07}\cite{LignierEtAl07,KierigEtAl08,DellaValleEtAl07}
of destruction of tunneling via time-periodic potential
differences~\cite{GrossmannEtAl91} was the breakthrough for
tunneling control via time-periodic potential differences. The
systems used so far in experiments are as divers as BECs in an
optical lattice~\cite{LignierEtAl07}, single particles in a double
well~\cite{KierigEtAl08} and light in a double-well
system~\cite{DellaValleEtAl07}.

Reference \cite{EckardtEtAl05} suggested to use time-periodic
shaking for a tilted double-well potential to measure an analog of
photon-assisted tunneling -- the ``photon''-frequencies
corresponding to shaking-frequencies in the kilo-Hertz regime.
Recently, photon-assisted tunneling was observed for a BEC in a
periodically shaken optical lattice~\cite{SiasEtAl08}. When the
BEC was allowed to expand for some time, the width of the
condensate was measured to be roughly proportional to the
tunneling matrix element for small BECs whereas it was
proportional to the square of the tunneling matrix element for
larger condensates.

While this might be an indication of a transition from ballistic
to diffusive transport~\cite{SiasEtAl08}, the verification of such
a transition requires the precise measurement of the
time-dependence of the width of the wave function in order to
distinguish both regimes of transport (see, e.g.,
Ref.~\cite{SteinigewegEtAl07} and references therein). Without
such measurements, other explanations for the observed dependence
of the width of the BEC cannot be excluded.

Here we develop an alternative interpretation of the experimental
results: Using two interacting bosons in a simple tight-binding
one-band model we show that the transition from a linear to a
quadratic dependence on the tunneling amplitude might be an
interaction-induced effect within the ballistic regime. We
demonstrate, both with the help of analytical arguments and
numerical simulations, that the wave function of an interacting
boson pair indeed behaves qualitatively very similar to the BEC in
the experiment: On the one hand, weakly interacting particles can
reproduce the linear scaling with the tunneling matrix element
observed for small condensates. On the other hand, more strongly
interacting particles reproduce the quadratic scaling for larger
condensates (which have a larger total interaction energy).

The paper is organized as follows. In Sec.~\ref{sec:mod} we
introduce the one-dimensional Bose-Hubbard Hamiltonian with a
constant force and time-periodic driving used to model the
experimental situation. In the case of high driving frequencies,
the time-dependent Hamiltonian can be replaced by an effective,
time-independent Hamiltonian. For this Hamiltonian, exact
two-particle energy eigenstates describing bound boson pairs can
be derived \cite{WinklerEtAl06,Jin08} as is explained in
Sec.~\ref{sec:exa}. In Sec.~\ref{sec:bal} we investigate the
dynamical behavior of two-particle wave packets. Employing the
obtained exact dispersion relation for interacting boson pairs, we
derive the dependence of the width of the wave packet after some
time of free expansion on the ratio of interaction and tunneling
matrix element. These theoretical considerations are supported by
numerical simulations of the two-particle Schr\"odinger equation
for the full driven Bose-Hubbard Hamiltonian. Some conclusions are
drawn in Sec.~\ref{conclu}.

\section{The Model\label{sec:mod}}
Using the notation of Ref.~\cite{SiasEtAl08}, the one-dimensional
Bose-Hubbard Hamiltonian with time-periodic shaking describing the
experiment can be written as:
\begin{eqnarray}
 \hat{H}_0 &=& -J\sum_j\left(\hat{c}^{\dag}_j\hat{c}^{\phantom{\dag}}_{j+1}
 +\hat{c}^{\dag}_{j+1}\hat{c}^{\phantom{\dag}}_j\right)
 + \frac U2\sum_j \hat{n}_j\left(\hat{n}_j-1\right)\nonumber\\
 & & +\Delta E \sum_jj\hat{n}_j + K\cos(\omega t)\sum_jj\hat{n}_j,
\label{eq:Htotal}
\end{eqnarray}
where the operators $\hat{c}^{(\dag)}_j$ annihilate (create)
bosons at the lattice site $j$ and $\hat{n}_j\equiv
\hat{c}^{\dag}_j\hat{c}^{\phantom{\dag}}_j$ are number operators;
$J$ is the hopping matrix element, $\Delta E \equiv Fd_L$ is the
potential difference of two adjacent wells with lattice spacing
$d_{\rm L}$, $\omega/(2\pi)$ the frequency with which the system
is shaken and $K$ the amplitude of the shaking. Thus, the
Hamiltonian (\ref{eq:Htotal}) describes a system of interacting
bosons in a tilted and driven optical lattice. For an untilted and
undriven lattice we have $\Delta E=0$ and $K=0$.

Floquet-theory~\nocite{Shirley65}\cite{Shirley65} can be applied
to understand the physics of such a driven system. For not too low
driving frequencies ($\hbar\omega\gg J$), the resonance condition
\begin{equation}
n\hbar\omega = \Delta E
\end{equation}
with integer $n$ leads to photon-assisted tunneling
(\cite{EckardtEtAl05,SiasEtAl08}). In the high-frequency limit
($\hbar\omega\gg J$ and $\hbar\omega\gg
U$~\cite{EckardtHolthaus07}) many aspects of the physics behind
the system can be understood by replacing the time-dependent
Hamiltonian with constant force by a time-independent Hamiltonian
without any additional force:
\begin{equation}
\hat{H}_{\rm eff} =
-J_{\rm eff}\sum_j\left(\hat{c}^{\dag}_j\hat{c}^{\phantom{\dag}}_{j+1}+\hat{c}^{\dag}_{j+1}\hat{c}^{\phantom{\dag}}_j\right)
+ \frac U2\sum_j \hat{n}_j\left(\hat{n}_j-1\right),
\label{eq:Heff}
\end{equation}
where the effective tunneling matrix element is given by
\begin{equation}
J_{\rm eff} = J{\cal J}_n(K_0),\quad K_0\equiv \frac K{\hbar \omega},
\label{eq:Jeff}
\end{equation}
with the $n$th order Bessel function ${\cal J}_n$. Thus, in this
time-periodic system photon assisted tunneling can lead to an
increase of the tunneling amplitude (as the energy difference
between neighboring wells is removed in the effective
Hamiltonian). Moreover, within an $n$-photon resonance, tunneling
can also be suppressed by tuning the driving amplitude $K$ such
that the ratio $K_0$  of driving amplitude and $\hbar \omega$
corresponds to a zero of the Bessel function~${\cal J}_n(K_0)$.

Without interactions ($U=0$), the effective Hamiltonian reduces to
a well-known single-particle Hamiltonian with extended Bloch-waves
as eigenfunctions
\begin{equation}
|\psi_k\rangle = \sum_{j=-\infty}^{\infty} e^{ikjd_{\rm L}}|j\rangle,
\label{eq:Bloch}
\end{equation}
where $|j\rangle$ is the Wannier-function at lattice-site $j$,
with the corresponding energy eigenvalues
\begin{equation}
E(k)=-2J_{\rm eff}\cos(kd_{\rm L}).
\label{eq:Ek}
\end{equation}
Within a parameter regime for which the $\cos$-dispersion relation
can be replaced by a quadratic dispersion relation, $E(k)\simeq
-2J_{\rm eff}+J_{\rm eff}k^2d_{\rm L}^2$, the single particle in a
tight-binding lattice behaves like a free particle. In fact, such
model Hamiltonians can be used to do numerics for a free particle
by setting
\begin{equation}
{J_{\rm eff}d_{\rm L}^2} = \frac{\hbar^2}{2m},
\label{eq:m}
\end{equation}
where $m$ is the mass of the free particle with dispersion
relation~$E_f(k)=\hbar^2k^2/(2m)$.

Without interaction, a BEC of $N$ bosons would simply be the
product of $N$ single-particle wave functions. If one measures the
width of a condensate after a certain time of free
expansion~\cite{SiasEtAl08}, this thus corresponds to the popular
text-book exercise~\cite{Fluegge90} of calculating the width
$\Delta x(t)$ for a single free particle. Starting from a Gaussian
wave packet at time zero,
\begin{equation}
 \psi(x,t=0) = \frac{1}{(2\pi a^2)^{1/4}}\exp\left(-\frac{x^2}{4a^2}\right),
\end{equation}
one finds
\begin{equation}
 \Delta x(t) = a\sqrt{1+\left(\frac{\hbar t}{2m a^2}\right)^2},
\end{equation}
where $\Delta x(t)^2 \equiv
\langle\psi(t)|x^2|\psi(t)\rangle-\langle\psi(t)|x|\psi(t)\rangle^2$.

For a non-interacting BEC in an optical lattice one can thus
expect to find for not too small free expansion times~$t$:
\begin{equation}
\Delta x(t) \propto \left| E''(0)\right|t,
\label{eq:deltax}
\end{equation}
where the dashes denote derivatives with respect to the argument
($k$). This equation was used in Ref.~\cite{SiasEtAl08} to measure
the effective tunneling matrix element $J_{\rm eff}$ as using
Eqs.~(\ref{eq:Ek}) and (\ref{eq:deltax}) one has:
\begin{equation}
\Delta x(t) \propto 2|J_{\rm eff}|d^2_{\rm L} t. \label{eq:dxjeff}
\end{equation}
In order to see if this relation always survives interaction, two
interacting bosons are investigated in the following sections,
starting with the construction of corresponding energy
eigenstates. Two-particle effects in optical lattices are interesting both experimentally and
theoretically~\cite{WinklerEtAl06}. Recent investigations include superexchange
interactions for mixtures of different spins~\cite{TrotzkyEtAl08}. Motivated by the
experiment~\cite{SiasEtAl08}, this manuscript concentrates
on pairs of indistinguishable particles.

\section{Exact analytic two-particle eigenfunctions\label{sec:exa}}
The aim is to find exact analytical expressions for the
eigenfunctions of the Hamiltonian~(\ref{eq:Heff}) for two bosons
and non-zero interaction $U\ne 0$. Rather than using the approach
via Green's functions of Ref.~\cite{WinklerEtAl06}, one can
proceed along the lines of Ref.~\cite{Weiss06b} to show in a
straightforward calculation (see the appendix for details) that a
large class of two-particle wave functions is given by
\begin{equation}
\label{eq:wellenfunkt}
|\phi_k\rangle = \sum_{\nu\le\mu} a_{\nu,\mu}(k)|\nu\rangle|\mu\rangle,
\end{equation}
where $\nu\le\mu$ is required because the bosons are
indistinguishable, and $|\nu\rangle|\nu\rangle$ corresponds to the
Fock state with two particles at lattice site $\nu$. The
coefficients $a_{\nu,\mu}$ are given by
\begin{equation}
\label{eq:koeffizienten}
a_{\nu,\mu}(k) = \left\{\begin{array}{lcl}
 b_{\nu,\mu}(k)&:&\mu\ne\nu,\\
b_{\nu,\mu}(k)/\sqrt{2}&:& \mu = \nu,
\end{array}\right.
\end{equation}
where
\begin{equation}
\label{eq:bdef}
b_{\nu,\mu}(k) = (\eta x_-)^{|\mu-\nu|}\exp\left[ikd_{\rm L}(\nu+\mu)\right]\;,
\end{equation}
\begin{equation}
\label{eq:etadef}
\eta = \left\{\begin{array}{lcl} -1&:& U/J_{\rm eff}>0,\\
+1&:& U/J_{\rm eff}<0,\end{array}\right.
\end{equation}
and
\begin{equation}
\label{eq:xmdef}
 x_-=\sqrt{\frac{U^2}{16J_{\rm eff}^2\cos^2(kd_{\rm L})}+1}-\frac{|U|}{|4J_{\rm eff}|\cos(kd_{\rm L})}\;.
\end{equation}
As for the Bloch-waves~(\ref{eq:Bloch}) and for plane waves for
free particles, these wave functions cannot be normalized in the
usual sense. Nevertheless, to avoid divergence for
$|\mu-\nu|\to\infty$, one needs $|x_-|\leq 1$ and thus $\cos(k
d_{\rm L})>0$.

The energy eigenvalue of the state $|\phi_k\rangle$ can be written
as (see Eq.~(\ref{eq:ap:en}) and Ref.~\cite{WinklerEtAl06}):
\begin{equation}
\label{eq:dispersion} E_2(k) = -4\eta J_{\rm
eff}\sqrt{\frac{U^2}{16J_{\rm eff}^2}+ \cos^2(kd_{\rm L})},
\end{equation}
and one thus obtains
\begin{equation}
\label{eq:e2} |E_2''(0)| =  \frac{16J_{\rm eff}^2d_{\rm
L}^2}{\sqrt{16J_{\rm eff}^2+U^2}}.
\end{equation}

The solutions calculated above represent the bound states of a
boson pair. In addition to these solutions one has a continuum of
scattering eigenstates \cite{Valiente08,InterinducedClass} which
will play, however, only a minor role in the description of the
physical situation considered here: The initial wave function is
given by a narrow Gaussian centered around a lattice site. Thus,
both particles are likely to be sitting at the same lattice site
and, therefore, the initial state lies nearly entirely in the
subspace of bound states of the two-boson system.

\section{Ballistic expansion of two-particle wave packets\label{sec:bal}}
The value that is relevant for the spreading of two-particle wave
packets is not the energy $E_2(k)$ of a pair but the energy per
particle. Hence, for the effective Hamiltonian (\ref{eq:Heff}) we
have
\begin{equation} \label{eq:e''}
 |E''(0)| =  \frac{8J_{\rm eff}^2d_{\rm L}^2}{\sqrt{16J_{\rm eff}^2+U^2}}.
\end{equation}
Equations (\ref{eq:deltax}) and (\ref{eq:e''}) therefore show that
the width of the BEC in the driven and tilted lattice is given by
\begin{equation} \label{eq:deltaxjeff}
 \Delta x_{J_{\rm eff}}(t) \propto \frac{8J^2_{\rm eff}d_{\rm L}t}{\sqrt{16J^2_{\rm eff}+U^2}},
\end{equation}
from which we find the following limiting behavior,
\begin{equation}
 \Delta x_{J_{\rm eff}}(t) \propto \left\{\begin{array}{rcl}2\left|J_{\rm eff}\right|
 d_{\rm L}^2t&:& \left|U/J_{\rm eff}\right|\ll 1,\\
 8J^2_{\rm eff}/|U| d_{\rm L}^2t&:& \left|U/J_{\rm eff}\right|\gg 1,
\end{array}\right.
\label{eq:dxend}
\end{equation}
where $J_{\rm eff}=J{\cal J}_n(K_0)$ was introduced in
Eq.~(\ref{eq:Jeff}). The corresponding expression for the width in
the undriven and untilted lattice, which we denote by $\Delta
x_J(t)$, is obtained by replacing $J_{\rm eff}$ by $J$,
\begin{equation} \label{eq:deltaxj}
 \Delta x_J(t) \propto \frac{8J^2d_{\rm L}t}{\sqrt{16J^2+U^2}}.
\end{equation}
Thus we see that the transition from a linear to a quadratic
dependence on the Bessel function observed in
Ref.~\cite{SiasEtAl08} can, within this simple two-particle model,
be explained as being a continuous transition based on the
dispersion relation (\ref{eq:dispersion}) for two interacting
particles.

It is important to note that one does not have to wait until the
dependence of $\Delta x(t)$ on $t$ becomes linear in order to see
the scaling when comparing, e.g., an undriven system without tilt
with a periodically shaken, tilted system with $n$-photon-assisted
tunneling. In the high-frequency limit, the main difference is
that the modulus of $J_{\rm eff}$ will be lower than $J$ by a
factor of $|{\cal J}_n(K_0)|$. If one plots the width of an
initially localized wave packet (cf.\ Ref.~\cite{SiasEtAl08}) as a
function of $\tau\equiv J t/\hbar$, the undriven system will thus
spread faster. Only by rescaling the time scale for the undriven
system one can hope to make both functions agree. For weak
interactions, the following equation relates the width of the
undriven system at time $|{\cal J}_n(K_0)|\tau$ to that of the
driven system at time $\tau$:
\begin{equation}
 \Delta x_J\left(|{\cal J}_n(K_0)|\tau\right) \simeq
 \Delta x_{J_{\rm eff}}(\tau)\;,\quad\left|\frac U{J_{\rm eff}}\right|\ll 1.
\end{equation}
For stronger interactions another factor of $|{\cal J}_n(K_0)|$ is
necessary:
\begin{equation} \label{eq:quadraticscaling}
 \Delta x_J\left(|{\cal J}_n(K_0)|^2\tau\right) \simeq
 \Delta x_{J_{\rm eff}}(\tau)\;,\quad\left|\frac U{J_{\rm eff}}\right|\gg 1.
\end{equation}

\begin{figure}
\includegraphics[width =1.2\linewidth,angle=-90]{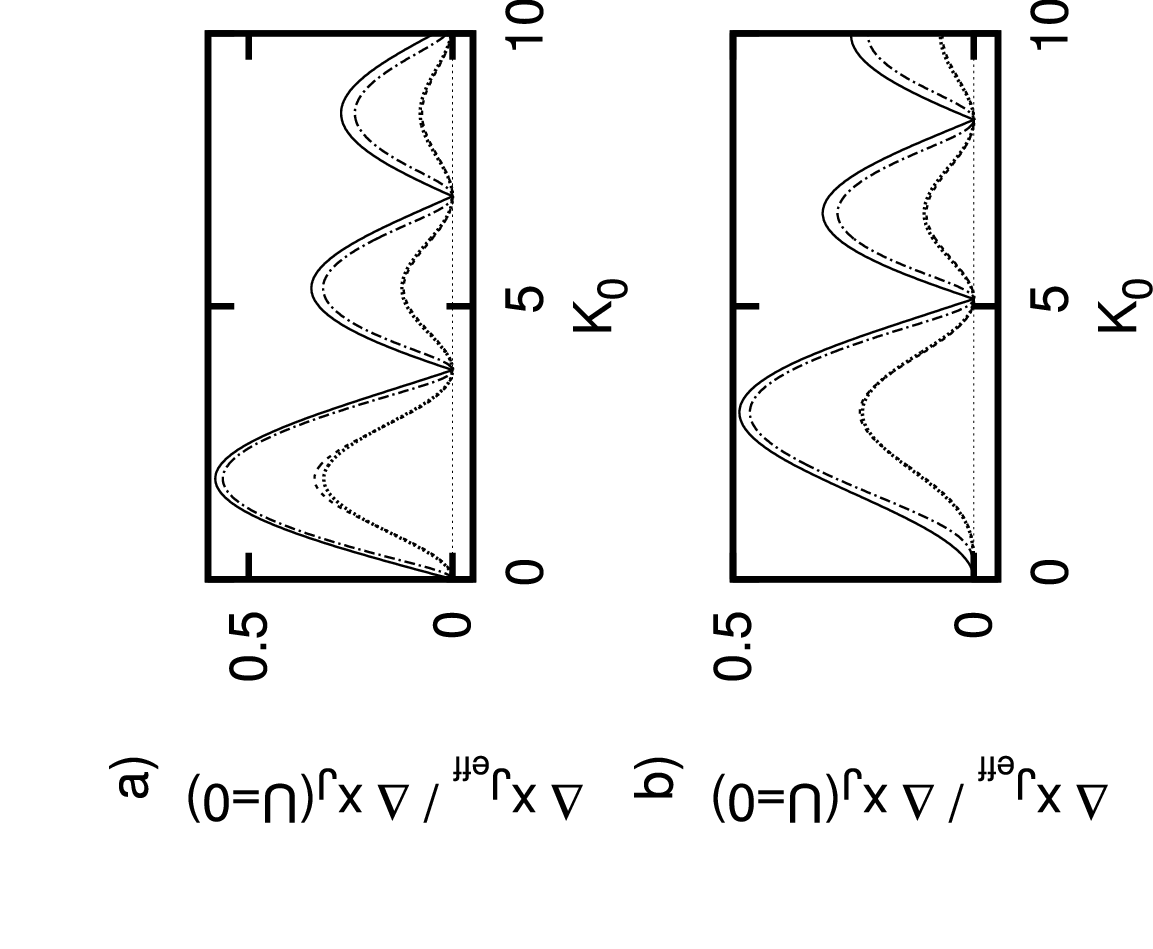}
\caption{The width of the two-particle wave function as a function
of the ratio of driving amplitude and driving frequency
$K_0=K/(\hbar \omega)$. The width is normalized to the value for
\textit{non}-interacting particles ($U=0$) in an undriven,
untilted lattice. The curves are based on the analytic
Eqs.~(\ref{eq:Jeff}), (\ref{eq:deltaxjeff}) and
(\ref{eq:deltaxj}). For the one-photon resonance (upper panel) one
has $J_{\rm eff}=J{\cal J}_1(K_0)$ and for the two-photon
resonance (lower panel) $J_{\rm eff}=J{\cal J}_2(K_0)$. Solid
curves: $|{\cal J}_n(K_0)|$, dashed curves: ${\cal J}^2_n(K_0)$,
dash-dotted curves: $U/J =0.6$, dotted curves: $U/J =3.6$. }
\label{fig:ana}
\end{figure}

\begin{figure}
\includegraphics[width =1.2\linewidth,angle=-90]{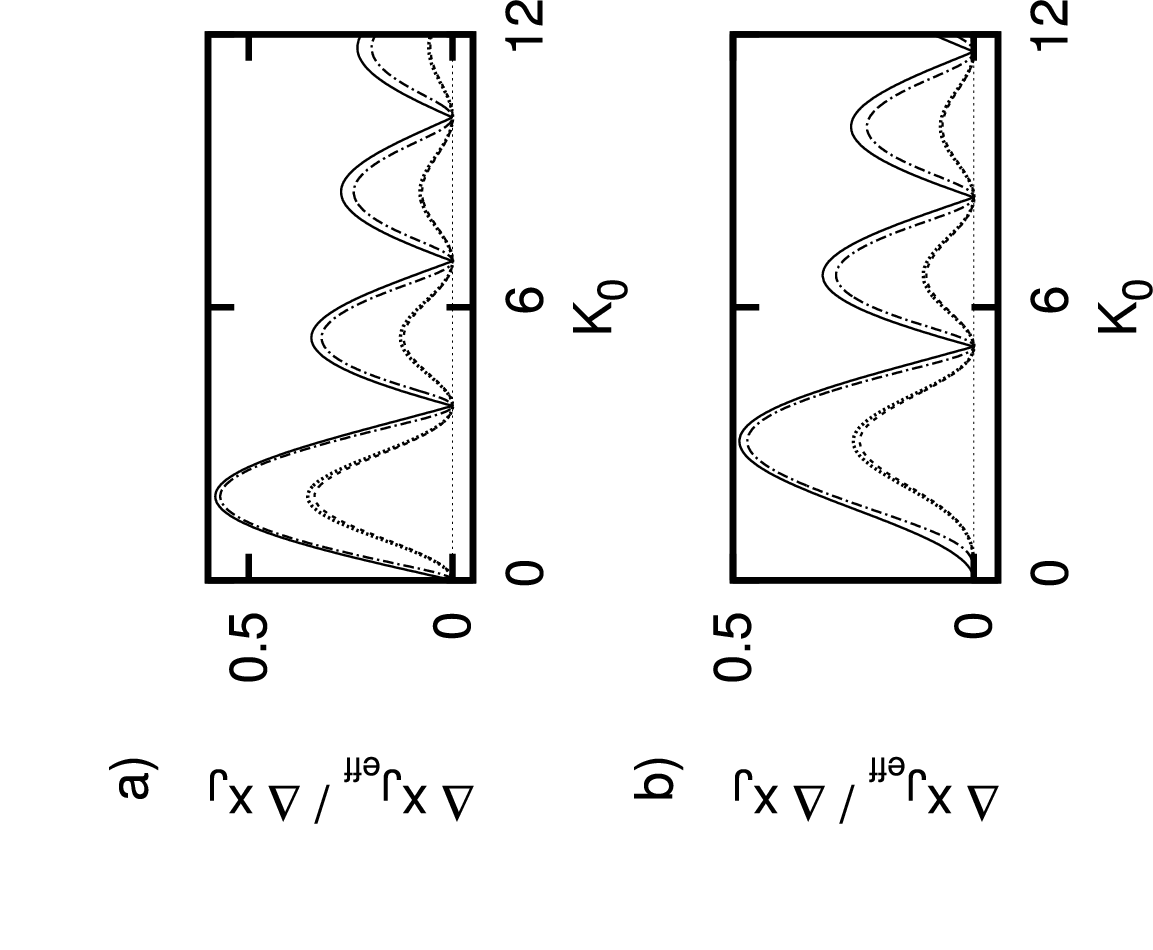}
\caption{The width  of the two-particle wave function as a
function of the ratio of driving amplitude and driving frequency
$K_0=K/(\hbar \omega)$ normalized to the value for
\textit{interacting}\/ particles  in an undriven, untilted
lattice.  Labels can be found in Fig.~\ref{fig:ana} but for the
dotted curves which are calculated for an interaction of $U/J=
10.0$.  As in Fig.~\ref{fig:ana}, the transition from linear to
quadratic dependence on ${\cal J}_n(K_0)$, $n=1,2$, can be
observed.} \label{fig:ana2}
\end{figure}

The above reasoning explains why the width of the BEC can be
proportional to ${\cal J}^2_n(K_0)$ even within the ballistic
regime. However, the experimental results of
Ref.~\cite{SiasEtAl08} show even more, namely that $\Delta
x_{J_{\rm eff}}(t)$ is not only proportional to ${\cal
J}^2_n(K_0)$, but also that the ratio of the width in the tilted
and driven lattice to the width in the untilted and undriven
lattice is approximately equal to ${\cal J}^2_n(K_0)$, i.~e.
\begin{equation} \label{eq:experiment}
 \frac{\Delta x_{J_{\rm eff}}(t)}{\Delta x_{J}(t)} \simeq
 {\cal J}^2_n(K_0).
\end{equation}
We demonstrate in Figs.~\ref{fig:ana} and \ref{fig:ana2} that it
is indeed possible to find parameters which reproduce the
experimentally observed behavior within the framework of the
theory presented here. In Fig.~\ref{fig:ana} we compare the driven
interacting system ($U\neq 0$) with the undriven system for
noninteracting particles ($U=0$). We see that for $U/J= 3.6$
(dotted curves) the left-hand side of Eq.~(\ref{eq:experiment})
nearly lies on top of the curves representing the right-hand side
of this equation (dashed curves). For lower interactions
(dash-dotted curves), the scaling is again $\propto |{\cal
J}_n(K_0)|$ as in the single-particle case.

In Fig.~\ref{fig:ana} the width of the wave function was compared
to the width for non-interacting particles. However, the observed
scaling even occurs (for larger interactions than those chosen in
Fig.~\ref{fig:ana}) when comparing the untilted, undriven
interacting system with the periodically driven interacting
system: Figure~\ref{fig:ana2} shows that again a very similar
scaling of the width of the condensate is found.

\begin{figure}
\includegraphics[width =1.0\linewidth,angle=-90]{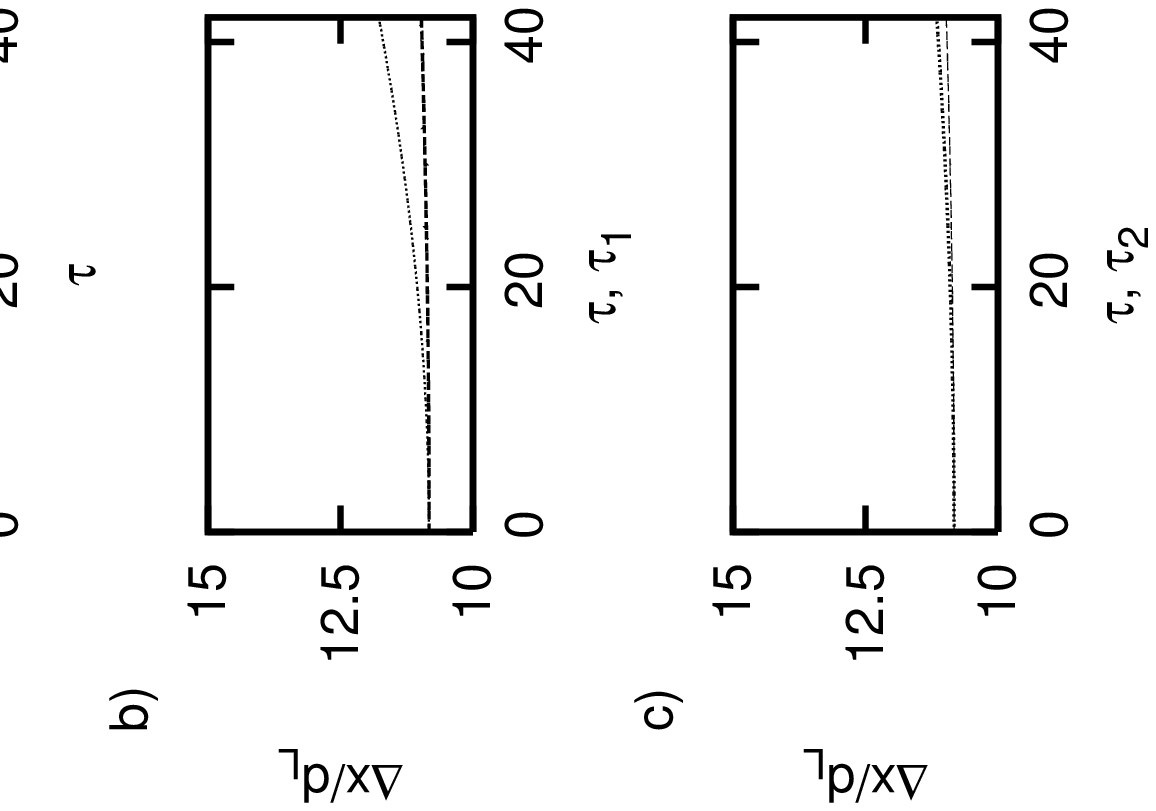}
\caption{Numeric simulation of a two-atom wave function in an
optical lattice with $N_{\rm L} =201$ lattice sites in the
high-frequency limit ($\hbar\omega=40J$) with large interactions
$U/J=10.0$ (cf.\ Fig.~\ref{fig:ana2}) and initial wave functions
given by Eq.~(\ref{eq:initial}). a) The width is plotted as a
function of dimensionless time $\tau\equiv Jt/\hbar$. Dotted
curves: the undriven, untilted lattice. Dashed curves: the same
lattice but with periodic driving for parameters corresponding to
a one-photon resonance~\cite{EckardtEtAl05, SiasEtAl08} ($K_0=2$,
cf.\ Eq.~(\ref{eq:Jeff})). The wave packet in the undriven system
spreads faster than in the driven system. b) Without the knowledge
of Figs.~\ref{fig:ana} and \ref{fig:ana2}, one might expect both
curves to agree if one plots $\Delta x(\tau_1)$, $\tau_1\equiv
|{\cal J}_1(2)|\tau$ for the undriven system. However, one needs
another factor of ${\cal J}_1(2)$: for $\tau_2\equiv [{\cal
J}_1(2)]^2\tau$, the data of the undriven system plotted as
$\Delta x(\tau_2)$ coincides with $\Delta x(\tau)$ for the driven
system (c).} \label{fig:num}
\end{figure}

Although several papers have shown the validity of the effective
Hamiltonian approach used so far (see, e.g.,
Ref.~\cite{EckardtHolthaus07} and references therein), one should
demonstrate that it also is valid for the present situation. To do
this, we numerically solve the time-dependent Schr\"odinger
equation corresponding to the full Hamiltonian~(\ref{eq:Htotal})
(Fig.~\ref{fig:num}). As the initial wave function we choose a
Gaussian function for the center-of-mass wave function
(corresponding to an initial confinement via a harmonic trapping
potential),
\begin{equation}
\label{eq:initial} |\psi(t=0)\rangle\equiv
\sum_{k}\exp\left(-a^2(k-k_0)^2/2\right)|\psi_k\rangle,
\end{equation}
where the sum over all possible $k$ values (rather than an
integral) is necessary as numeric calculations cannot be done in
infinite lattices. The initial wave function is sitting in the
middle of the lattice (for an odd number $N_{\rm L}$ of
lattice-sites, the sites can be labelled as \mbox{$j=-(N_{\rm
L}-1)/2\ldots (N_{\rm L}-1)/2$}). In the finite, shaken lattice
relevant for the numerics in this paper, vanishing boundary
conditions are a suitable choice~\cite{vanishing}.  For the
center-of-mass part of the wave function (cf.\
Eq.~(\ref{eq:koeffizienten})), $\mu_{\rm c.o.m.}=(\mu+\nu)/2$, the
wave-vector is $2 k$, possible values for $k$ are thus
$n\pi/[(N_{\rm L}-1)d_{\rm L}], n=1,2,3,\ldots$. The initial wave
function was calculated without any initial momentum ($k_0=0$) and
with $a=10d_{\rm L}$. For the undriven system, $U/J=10.0$ was
chosen in the initial wave function, for the driven system with
$K_0=2.0$, we chose $U/J=10.0/{\cal J}_1(2.0)$ to mimic an
experimental situation where the initial wave function is prepared
in a harmonic-oscillator potential and the periodic shaking is
switched on before switching off this potential. The quadratic
scaling as predicted in Eq.~(\ref{eq:quadraticscaling}) can thus
indeed be observed numerically (Fig.~\ref{fig:num}).

\section{Conclusion}\label{conclu}
Photon-assisted tunneling in a periodically shaken optical lattice
was investigated for two interacting bosons. Both numerical and
analytical calculations were done for a periodically driven
one-band Bose-Hubbard model. Figures~\ref{fig:ana}-\ref{fig:num}
demonstrate that the experimentally observed~\cite{SiasEtAl08}
dependence of the width of the BEC on the square of the tunneling
matrix element can be explained -- at least qualitatively within
the simple two-particle model investigated here -- as being an
interaction-induced effect which is based on the dispersion
relation for bound boson pairs.

While our simplified approach thus explains some aspects of the
experiment~\cite{SiasEtAl08}, calculations for larger particle
numbers (which are not straightforward to generalize from the
method presented here) are likely to lead to further insights into
the experiment for which additional effects like decoherence by
particle losses might also play an important role. An experimental
measurement of the time-dependence of the width of the wave
function would be of great interest for the theoretical analysis
and the modelling of the transport properties of quantum
condensates. In particular, it would be interesting to see if
indeed a transition from ballistic to diffusive transport takes
place, or if the simple two-particle model presented in this paper
can explain the relevant features of the experiment.

\acknowledgments

We would like to thank Y.~Castin, A.~Eckardt, M.~Holthaus, O.~Morsch and
N.~Teichmann for insightful discussions. CW gratefully
acknowledges funding by the EU (contract MEIF-CT-2006-038407). HPB
gratefully acknowledges financial support within a fellowship of
the Hanse-Wissenschaftskolleg, Delmenhorst.

\begin{appendix}

\section{Eigenfunctions}
In order to show that exact two-particle
eigenfunctions~\cite{InterinducedClass} of the effective
Hamiltonian~(\ref{eq:Heff}) are indeed given by
Eqs.~(\ref{eq:wellenfunkt}) and (\ref{eq:koeffizienten}), one can
start with
\begin{eqnarray}
\sum_j\hat{c}^{\dag}_j\hat{c}^{\phantom{\dag}}_{j+1}|\mu\rangle|\mu\rangle &=&
\sqrt{2}|\mu-1\rangle|\mu\rangle\;,\\\nonumber
\sum_j\hat{c}^{{\dag}}_{j+1}\hat{c}^{\phantom{\dag}}_j|\mu\rangle|\mu\rangle &=& \sqrt{2}|\mu\rangle|\mu+1\rangle\;,\\\nonumber
\sum_j\hat{c}^{\dag}_j\hat{c}^{\phantom{\dag}}_{j+1}|\mu-1\rangle|\mu\rangle &=&
|\mu-2\rangle|\mu\rangle+\sqrt{2}|\mu-1\rangle|\mu-1\rangle\;,\\\nonumber
\sum_j\hat{c}^{\dag}_{j+1}\hat{c}^{\phantom{\dag}}_{j}|\mu-1\rangle|\mu\rangle &=& |\mu-1\rangle|\mu+1\rangle+\sqrt{2}|\mu\rangle|\mu\rangle\;,
\end{eqnarray}
and for $\nu<\mu-1$
\begin{eqnarray}
\sum_j\hat{c}^{\dag}_j\hat{c}^{\phantom{\dag}}_{j+1}|\nu\rangle|\mu\rangle =
|\nu-1\rangle|\mu\rangle+|\nu\rangle|\mu-1\rangle\;,\\ \nonumber
\sum_j\hat{c}^{\dag}_{j+1}\hat{c}^{\phantom{\dag}}_{j}|\nu\rangle|\mu\rangle =
|\nu+1\rangle|\mu\rangle+|\nu\rangle|\mu+1\rangle\;.
\end{eqnarray}
Using the notation of Eq.~(\ref{eq:wellenfunkt})
one thus has
\begin{eqnarray}
\label{eq:neben1}
\hat{H}_{\rm eff}|\phi_k\rangle=
 U\sum_{\mu} a_{\mu, \mu}(k)|\mu\rangle|\mu\rangle \quad\quad\quad\quad\quad\quad\\\nonumber
- J_{\rm eff}\sum_{\mu}\sqrt{2}a_{\mu,
  \mu}(k)\left(|\mu-1\rangle|\mu\rangle+|\mu\rangle|\mu+1\rangle\right)\\\nonumber
- J_{\rm eff}\sum_{\mu} a_{\mu-1,
  \mu}(k)\left(|\mu-2\rangle|\mu\rangle+\sqrt{2}|\mu-1\rangle|\mu-1\rangle\right)\\\nonumber
- J_{\rm eff}\sum_{\mu} a_{\mu-1,
  \mu}(k)\left(|\mu-1\rangle|\mu+1\rangle+\sqrt{2}|\mu\rangle|\mu\rangle\right)\\\nonumber
- J_{\rm eff}\sum_{\nu<\mu-1}
a_{\nu,\mu}(k)\left(|\nu-1\rangle|\mu\rangle+|\nu\rangle|\mu-1\rangle\right)\\\nonumber
- J_{\rm eff}\sum_{\nu<\mu-1}
a_{\nu,\mu}(k)\left(|\nu+1\rangle|\mu\rangle+|\nu\rangle|\mu+1\rangle\right).
\end{eqnarray}
In oder to show that this indeed leads to an eigenfunction of the
effective Hamiltonian~(\ref{eq:Heff}), we use
Eq.~(\ref{eq:koeffizienten}) and start with the last two lines of
Eq.~(\ref{eq:neben1}). Performing appropriate shifts of the
summation indices these lines can be written as
\begin{eqnarray} \label{term-1}
 &-& J_{\rm eff}\sum_{\nu<\mu}\left(
 b_{\nu,\mu+1}(k) + b_{\nu-1,\mu}(k) \right) |\nu\rangle|\mu\rangle
 \nonumber \\
 &-& J_{\rm eff}\sum_{\nu<\mu-2}\left(
 b_{\nu+1,\mu}(k) + b_{\nu,\mu-1}(k) \right) |\nu\rangle|\mu\rangle.
\end{eqnarray}
The second, third and fourth line of Eq.~(\ref{eq:neben1}) can be
combined to yield
\begin{eqnarray} \label{term-2}
 &-& J_{\rm eff}\sum_{\nu}\sqrt{2}\left(
 b_{\nu-1,\nu}(k) + b_{\nu,\nu+1}(k) \right) |\nu\rangle|\nu\rangle
 \nonumber \\
 &-& J_{\rm eff}\sum_{\nu=\mu-1}\left(
 b_{\nu+1,\mu}(k) + b_{\nu,\mu-1}(k) \right) |\nu\rangle|\mu\rangle
 \nonumber \\
 &-& J_{\rm eff}\sum_{\nu=\mu-2}\left(
 b_{\nu+1,\mu}(k) + b_{\nu,\mu-1}(k) \right) |\nu\rangle|\mu\rangle.
\end{eqnarray}
Adding \eqref{term-1} and \eqref{term-2} and including the first
line of Eq.~\eqref{eq:neben1} we find
\begin{eqnarray} \label{eq:neben2}
 \hat{H}_{\rm eff}|\phi_k\rangle &=&
 \frac{U}{\sqrt{2}}\sum_{\nu} b_{\nu, \nu}(k)|\nu\rangle|\nu\rangle
 \quad\quad\quad\quad\quad\quad \\\nonumber
 &-& J_{\rm eff}\sum_{\nu}\sqrt{2}\left(
 b_{\nu-1,\nu}(k) + b_{\nu,\nu+1}(k) \right) |\nu\rangle|\nu\rangle
 \nonumber \\
 &-& J_{\rm eff}\sum_{\nu<\mu}\left(
 b_{\nu,\mu+1}(k) + b_{\nu-1,\mu}(k) \right) |\nu\rangle|\mu\rangle
 \nonumber \\
 &-& J_{\rm eff}\sum_{\nu<\mu}\left(
 b_{\nu+1,\mu}(k) + b_{\nu,\mu-1}(k) \right) |\nu\rangle|\mu\rangle.
 \nonumber
\end{eqnarray}
In order to simplify the last two lines of Eq.~(\ref{eq:neben2})
we use the notation of Eqs.~(\ref{eq:bdef})-(\ref{eq:xmdef}) to
obtain
\begin{eqnarray*}
 \lefteqn{
 b_{\nu+1,\mu}(k)+b_{\nu-1,\mu}(k) + b_{\nu,\mu+1}(k)+b_{\nu,\mu-1}(k) } \\
 &=&  b_{\nu,\mu}(k)\eta \Big[ x_-^{-1}e^{ikd_{\rm L}}+x_-e^{-ikd_{\rm L}} \\
 && \qquad\qquad +x_-e^{ikd_{\rm L}} +x_-^{-1}e^{-ikd_{\rm L}}\Big] \\
 &=& b_{\nu,\mu}(k)2\eta \cos\left(kd_{\rm L}\right)\left[x_-+x_-^{-1}\right]
\end{eqnarray*}
with
\begin{eqnarray*}
 x_-+x_-^{-1} =2\left(\frac{U^2}{16J_{\rm eff}^2\cos^2(kd_{\rm L})}+1\right)^{1/2}.
\end{eqnarray*}
Thus, for $\nu<\mu$ we have
\begin{eqnarray}
 \lefteqn{
 \langle\mu|\langle\nu|\hat{H}_{\rm eff}|\phi_k\rangle } \\
 &=&-4\eta J_{\rm eff}\cos(kd_{\rm L})\left(\textstyle{\frac{U^2}{16J_{\rm eff}^2\cos^2(kd_{\rm
      L})}+1}\right)^{1/2} a_{\nu,\mu}(k). \nonumber
\end{eqnarray}
The terms with $\nu=\mu$ in Eq.~(\ref{eq:neben2}) yield:
\begin{eqnarray*}
 \lefteqn{
 (U/\sqrt{2})b_{\nu,\nu}(k) -J_{\rm eff}\sqrt{2}\left[b_{\nu,\nu+1}(k)+b_{\nu-1,\nu}(k)\right] }\\
 &=& \left[U/\sqrt{2}-J_{\rm eff}\sqrt{2}x_-2\eta \cos(kd_{\rm L})\right]b_{\nu,\nu}(k) \\
 &=& \left[U-J_{\rm eff}4x_-\eta \cos(kd_{\rm L})\right]a_{\nu,\nu}(k),
\end{eqnarray*}
and thus (cf.\ Eq.~(\ref{eq:xmdef})):
\begin{eqnarray}
 \lefteqn{
 \langle\nu|\langle\nu|\hat{H}_{\rm eff}|\phi_k\rangle } \nonumber \\
 &=& -4\eta J_{\rm eff}\cos(kd_{\rm L})\left({\textstyle\frac{U^2}{16J_{\rm eff}^2\cos^2(kd_{\rm
      L})}+1}\right)^{1/2} a_{\nu,\nu}(k) \nonumber \\
 &&+\left[U+\eta J_{\rm eff}\frac{|U|}{|J_{\rm eff}|}\right]
 a_{\nu,\nu}(k),
\end{eqnarray}
where $U+\eta J_{\rm eff}{|U|}/{|J_{\rm eff}|}=0$
[Eq.~(\ref{eq:etadef})]. This shows that $|\phi_k\rangle $ indeed
is a 2-particle eigenfunction of the effective Hamiltonian,
\[
 \hat{H}_{\rm eff}|\phi_k\rangle = E_2(k)|\phi_k\rangle,
\]
with the energy eigenvalue
\begin{equation}
\label{eq:ap:en} E_2(k) = -4\eta J_{\rm
eff}\sqrt{\frac{U^2}{16J_{\rm eff}^2}+ \cos^2(kd_{\rm L})}.
\end{equation}
Note that $\cos(kd_{\rm L})$ was required to be positive for $x_-$
to have a modulus lower than or equal to one, because otherwise
the wavefunction would diverge.
\end{appendix}


\begin{thebibliography}{27}
\bibitem[{van()}]{vanishing}
\bibinfo{note}{Instead of $\exp\left[ikd_{\rm L}(\nu+\mu)\right]$, the center
  of mass wave-functions in Eqs.~(\ref{eq:initial}) and
  (\ref{eq:koeffizienten}) alternately are given by $\cos\left[ikd_{\rm
  L}(\nu+\mu)\right]$ and $i\sin\left[ikd_{\rm L}(\nu+\mu)\right]$.}
\end{thebibliography}

\end{document}